\documentclass[nocite]{epl}
\usepackage{amssymb}
\usepackage{graphicx}
\usepackage{amssymb}
\usepackage{epstopdf}

\title{Flow-injection of branched polymers inside nanopores}
\author{Takahiro Sakaue\inst{1} \and Elie Rapha\"el\inst{1} \and Pierre-Gilles de Gennes\inst{2} \and Fran\c{c}oise Brochard-Wyart\inst{2}}
\institute{
  \inst{1} Physique de la Mati\`ere Condens\'ee and Mati\`ere et Syst\`emes Complexes, UMR 7057 CNRS, Coll\`ege de France, 11 place Marcelin Berthelot, 75231 Paris Cedex 05, France\\
  \inst{2} Institut Curie, 11 rue Pierre et Marie Curie, 75231 Paris Cedex 05, France
  }

\begin{document}

\maketitle

\begin{abstract}
Flexible chains (linear or branched) can be forced to enter into a narrow capillary by using
a hydrodynamic flow.
Here, we correct our earlier description of this problem by considering the progressive nature of the suction process.
We find that the critical current for penetration, $J_c$,  is controlled by the entry of a single blob of the capillary size, and that its scaling structure is the same for branched and linear chains.
\end{abstract}

\section{Introduction}
Confined polymer chains show up in many sectors ranging from filtration problems to DNA permeation~\cite{nanopore}.
In good solvent conditions, the concentration of linear chains in a pore smaller than the coil size, in equilibrium with a dilute bulk solution, is exponentially small~\cite{deGennes}.
However, the chain may be sucked in the pore by a flow (with an overall current $J$) if the current is beyond a certain threshold given
 by~\cite{deGennes_review}:
\begin{eqnarray}
J_c \simeq k_BT/\eta
\label{J_c_linear}
\end{eqnarray}
where $\eta$ is the solvent viscosity and $k_BT$ is the thermal energy.
Note that this critical current is independent of both the molecular weight of the chains and the capillary size.

Ten years ago, some of us tried to extend this discussion to the case of branched chains~\cite{deGennes_review,Gay}.
The hope was that the threshold current could bring new informations on the properties of branched objects. It turns out, however, that our discussion was incorrect!
The purpose of the present letter is to improve on that.

The incorrect result was based on an  ``inside approach", where the confined chain is considered independently of the entry.
The implicit assumption was that the structure of a partly sucked chain may be represented by that of a completely confined chain.

Below, after recalling some basic conformational principles of branched polymers, we summarize the earlier argument and show its weakness.
We then present a Flory type argument for a partly sucked branched chain and discuss the progressive nature of the confining process.

\section{Statics of branched polymers}
We start with a branched polymer made up of $N$ monomers of size $a$, in solution in a good solvent.
We assume the molecule to be flexible, and characterized by a tree-like structure with no loops.
The ideal size ({\it{i.e.}} the size without volume interactions) of such an object is given by 
$R_0 \simeq a N^{1/4}$~\cite{branch_Zimm}.
The chain swelling due to excluded volume interactions can be derived using a Flory type argument as follows~\cite{Isaacson, Daoud_Joanny}.
The free energy of the branched chain with size $R$ is
\begin{eqnarray}
\frac{F}{k_BT} \simeq \frac{R^2}{R_{0}^2}+ \frac{N^2a^3}{R^3}
\label{Flory1}
\end{eqnarray}
The first term is an elastic term, and the second term comes from the intermonomer repulsions.
Minimizing Eq.(\ref{Flory1}) with respect to $R$ yields the equilibrium size of the chain:
\begin{eqnarray}
R\simeq a N^{1/2}
\label{R_branch}
\end{eqnarray}
When such a branched object is confined into a capillary of size $D < R$,
its conformational properties are significantlly altered~\cite{Vilgis_1} and
the chain is extended over a length $L$ along the capillary axis.
The Flory free energy of the chain is now given by:
\begin{eqnarray}
\frac{F}{k_BT} \simeq \frac{L^2}{R_{0}^2}+ \frac{N^2a^3}{LD^2}
\label{flory2}
\end{eqnarray}
Minimizing Eq.(\ref{flory2}) with respect to $L$ yields:
\begin{eqnarray}
\frac{L}{D} \simeq \left( \frac{a}{D}\right)^{5/3} N^{5/6}
\label{y_part}
\end{eqnarray}
The confined branched object is reminiscent of semidilute solutions and can be seen as a compact stacking of blobs of size $\xi$ with a number of monomers $g$~\cite{Gay,deGennes_review}.
Inside a blob, the effect of confinement is negligible, and $\xi$ and $g$ are related by
$\xi = a g^{1/2}$ (see Eq.(\ref{R_branch})).
Since the blobs are compactly stacked, $\xi$ and $g$
must also satisfy the relation 
$\phi \simeq g a^3/\xi^3$, where $\phi$ is the volume fraction
given by $\phi \simeq Na^3/LD^2$.
From the above equations we obtain for the mesh size
\begin{eqnarray}
\xi \simeq  a \left(\frac{D}{a} \right)^{4/3}N^{-1/6}  \simeq D \left( \frac{D}{L} \right)^{1/5}
\label{xi}
\end{eqnarray}
For $D \simeq R$, we have $\xi \simeq R$. Note also that $\phi \simeq 1$,
 $\xi \simeq a$ and $L \simeq a N^{3/4}$ for $D \simeq a N^{1/8}$ (this latter value
 is called the {\it minimum diameter}~\cite{Vilgis_1}).

\section{Earlier approach}
\begin{figure}
\onefigure{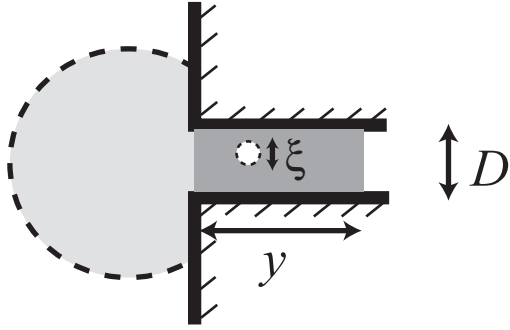}
\caption{A partly injected polymer inside a narrow capillary. A part inside the capillary is a dense piling of blobs of size $\xi$.}
\label{process_old}
\end{figure}

In this section, we briefly review the earlier "inside approach" for the flow injection of branched polymers~\cite{deGennes_review,Gay,star}.
Let us consider the situation in which part of the polymer chain is located inside a pore of diameter $D$ (with $D < a N^{1/2}$, see Fig. \ref{process_old})). 
\label{previous_approach}
The length of the pore occupied by the chain is denoted by $y$.
The  free energy, $F(y)$, associated with  the confined part of the chain is given by
\begin{eqnarray}
F(y) = f_{{\rm conf}}  \ y - \int_0^y f_{{\rm hyd}}(y) dy
\label{f_y}
\end{eqnarray}
In the above equation, the force due to the confinement is
\begin{eqnarray}
f_{{\rm conf}} \simeq \frac{k_BT}{\xi^3} D^2
\label{f_conf}
\end{eqnarray}
The hydrodynamic force $f_{{\rm hyd}}$ can be estimated as a Stokes drag per blob.
Assuming that $y$ is larger than the blob size, we get
\begin{eqnarray}
f_{{\rm hyd}} \, \simeq {\eta \, \xi \, v(D)} \; \frac{D^2 y}{\xi^3}
\label{f_hydro}
\end{eqnarray}
where $v(D) = J/(\pi D^2)$ is the local solvent velocity inside the capillary and $(D^2 y)/{\xi^3}$ is the number of blobs injected.
The free energy Eq.(\ref{f_y}) can thus be rewritten as
\begin{eqnarray}
F(y) \simeq \frac{k_BT}{\xi^3} D^2 y -\frac{1}{2}\frac{\eta J}{\xi^2}y^2
\label{f_y_order}
\end{eqnarray}
If one assumes that $\xi$ is identical to the equilibrium mesh size of a chain entirely confined
into the capillary, then $\xi$ is independent of $y$ and Eq.(\ref{f_y_order}) indicates
the existence of a barrier in the free energy 
\begin{eqnarray}
F^* \simeq \frac{(k_BT)^2}{\eta J} \left( \frac{D}{\xi}\right)^4
\label{F_barrier}
\end{eqnarray}
located at 
\begin{eqnarray}
y = y^* \simeq \frac{k_BT}{\eta J} \frac{D^2}{\xi}
\label{y_barrier}
\end{eqnarray}
If $F^*$ is smaller than the thermal energy $k_B T$, 
the injection process takes place.
Therefore, the critical current $J_c$ for a branched polymer is predicted to be 
given by~\cite{Gay}:
\begin{eqnarray}
\frac{\eta}{k_BT} J_c \simeq \left( \frac{D}{\xi} \right)^4
\label{J_c_previous}
\end{eqnarray}
Using Eq.(\ref{xi}), Eq.(\ref{J_c_previous}) can be rewritten as
\begin{eqnarray}
\frac{\eta}{k_BT} J_c \simeq \left( \frac{a}{D}\right)^{4/3} N^{2/3}
\label{J_c_previous_branch}
\end{eqnarray}
which indicates that the larger branched polymer would be injected into smaller pore with more difficulties.

Equation ~(\ref{J_c_previous}), (\ref{J_c_previous_branch}) were the main predictions of ref.~\cite{Gay}, but they are incorrect for the following reason. The threshold length $y^*$ corresponding  to $J = J_c$ is given by (see Eqs.(\ref{y_barrier},\ref{J_c_previous}))
\begin{eqnarray}
y^*_{J=J_c} \simeq \frac{\xi^3}{D^2}
\label{y_c_J_c}
\end{eqnarray}
Equation.(\ref{y_c_J_c}) indicates that $y^*_{J=J_c} < \xi$.
But at $y < \xi$, the polymer does not feel yet 
the effect of the confinement at all ! This contradiction clearly indicates that the assumption that $\xi$ is identical to the equilibrium mesh size of an entirely confined chain is incorrect.

As we will now show, the mesh size $\xi$ depends in fact on the degree of penetration, $y$, and can be calculated based on a Flory method for a {\it finite} piece of chain.

\section{Flory approach for a partly sucked branched chain}
We now focus our attention to the suction process.
The statistical properties of a partly sucked branched object can be discribed based on a Flory type of calculation.

The procedure is the same as the case for the completely confined chain, except for the point that now only a part of the chain (with the number of monomers $P <N$) is inside the capillary.
The sucked part of the chain is extended over a length $y$ along the tube and its degree of extension can be derived from the optimization of the following free energy.
\begin{eqnarray}
\frac{F}{k_BT} \simeq \frac{y^2}{R_{0(P)}^2}+ \frac{P^2a^3}{yD^2}
\label{Flory2}
\end{eqnarray}
where $R_{0(P)}^2 = a P^{1/4}$ is the size of an ideal branched chain made from $P$ monomers.
Minimizing eq.~(\ref{Flory2} ) with respect to $y$, we obtain
\begin{eqnarray}
\frac{y}{D} \simeq \left( \frac{a}{D}\right)^{5/3} P^{5/6}
\label{y_part}
\end{eqnarray}
The correlation length for the confined section is calculated in an analogious way.
For a partly sucked branched chain, we obtain the $y$ dependent mesh size
\begin{eqnarray}
\xi(y) \simeq D \left( \frac{D}{y} \right)^{1/5}
\label{xi_y}
\end{eqnarray}
All these results, of course, coincide with the case of completely confined chain by replacing $P$ with $N$.
It is important to note that the mesh size $\xi (y)$ is a decreasing function of $y$.
This means that the branched object is progressively compressed during the confinement process (Fig.~\ref{branch_tube}).

With progressive confining in mind, we now return to our discussion of suction process.
We may still use the ``inside" discription of eq.~(\ref{f_y})-(\ref{f_hydro}).
Using the $y$ dependent mesh size eq.~(\ref{xi_y}), we arrive at
\begin{eqnarray}
F(y) \simeq k_BT \left( \frac{y}{D} \right)^{8/5} -\eta J \left( \frac{y}{D} \right)^{12/5}
\end{eqnarray}
The above free energy display an energy barrier of height $F^*$, located at $y = y^*$, with
\begin{eqnarray}
F^* \simeq k_BT \left( \frac{y^*}{D} \right)^{8/5}
\end{eqnarray}
and\begin{eqnarray}
y^* \simeq (k_BT/(\eta J))^{5/4} D
\label{y_barrier_new}
\end{eqnarray}
At the threshold, $F^* \simeq k_BT$, therefore we arrive at a remarkable result
\begin{eqnarray}
y^*_{J=J_c} \simeq D = \xi(y\simeq D)
\label{y_barrier_Jc}
\end{eqnarray}
In contrast to Eq.(\ref{y_c_J_c}), the corrected value of $y^*$ (Eq.(\ref{y_barrier_Jc}))
is now comparable to the value of $\xi$ at the entrance of the pore and eq.~(\ref{f_y})-(\ref{f_hydro}) are therefore meaningful.
From eq.~(\ref{y_barrier_new}) and (\ref{y_barrier_Jc}), we find that the critical current $J_c$ for branched chains is still given by the scaling form of eq.~(\ref{J_c_linear}): we expect no dramatic difference between branched chains and linear chains.

\section{Discussion}
We have shown that the scaling structure of the critical current for penetration, $J_c$, 
is the same for branched and linear chains.
Though suprizing, this result can be explained physically.
Our argument indicates that the crucial moment is the insertion of a {\it single blob} of the capillary size.
We might ask whether more resistance comes out from the second, third blob...
\begin{figure}
\onefigure{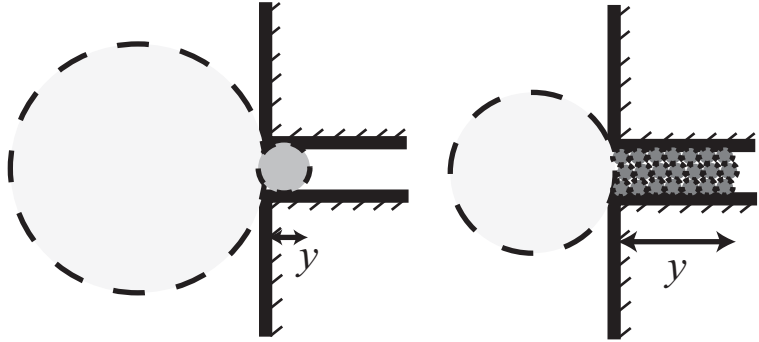}
\caption{The suction process of a branched polymer into a capillary. The colour strength represents the density.}
\label{branch_tube}
\end{figure}
The crucial physical feature is a runaway: starting from one blob, if we go to higher value of $y$, the hydrodynamic drag increases faster with $y$ than the confinement force.
This can be checked in detail starting from Eqs.~(\ref{f_y} - \ref{f_hydro}),  and allowing $\xi_{in}$ to depend on $y$ (eq.~(\ref{xi_y})).
Such a feature is associated with the progressive nature of the confining process of branched objects into the capillary (Fig.~\ref{branch_tube})\footnote{We note that a branched polymer in a capillary is not an exceptional case. As a simple example, a linear chain in a closed cavity shows physically same behaviours. The corresponding confinement regime may be called as {\it strong confinement} and should be distinguished from {\it weak confinement} regime, where the mesh size does not depend on the degree of confinement~\cite{confine}.}.
Thus, as soon as one blob has entered the capillary, the whole structure should get sucked in.

From a practical point of view, we now expect that $J_c$ is not sensitive to branching (except for numerical coefficients which go beyond scaling).
Our hope of probing branched structure by suction experiments is thus destroyed.
However, for more complicated structures (stars, nanogels), the suction process may be interesting.
This is discussed in a separated paper by one of us (T.S.).



\acknowledgments
T.S. was supported by a Research Fellowship from the Japan Society for the Promotion of Science for Young Scientists (No. 4990).

\end{document}